\documentclass[aps,prb,onecolumn,showpacs,groupedaddress]{revtex4-1} 
\usepackage{graphicx}  
\usepackage{dcolumn}   
\usepackage{bm}        
\usepackage{amssymb}   
\usepackage{times}
\begin{document}
\title{Importance of non-parabolic band effects in the thermoelectric
properties of semiconductors}

\author{Xin Chen}
\author{David Parker}
\author{David J. Singh}

\affiliation{Materials Science and Technology Division,
Oak Ridge National Laboratory, Oak Ridge, TN 37831-6056}
\date{\today}

\begin{abstract}
We present an analysis of the thermoelectric properties of
of $n$-type GeTe and SnTe in relation to the
lead chalcogenides PbTe and PbSe.
We find that the singly degenerate conduction bands
of semiconducting GeTe and SnTe are
highly non-ellipsoidal, even very close to the band edges.
This leads to isoenergy surfaces
with a strongly corrugated shape that is clearly
evident at carrier concentrations well below 0.005 $e$ per formula unit (7 - 9 $\times$ 10$^{19}$cm$^{-3}$ depending on material)
Analysis within Boltzmann theory
suggests that this corrugation may be
favorable for the thermoelectric transport.
Our calculations also indicate that values of the power factor
for these two materials may well exceed 
those of PbTe and PbSe.
As a result these materials may exhibit $n$-type
performance exceeding that of the lead chalcogenides.
\end{abstract}

\maketitle

Electronic properties of
semiconductors, and thermoelectrics in particular, are typically 
understood using parabolic band models.
Within these models the electronic structure and transport
is governed by an effective mass, and anisotropy
is quantified by the use of anisotropic effective mass tensors.
Non-parabolicity in finite gap
semiconductors is normally characterized by an energy
and/or temperature dependent mass, and is taken as a weak effect
as it must be sufficiently close to the band edge with a finite gap.
Transport is then understood in terms of parabolic band expressions
with a varying effective mass.
In the context of thermoelectrics these expressions imply
limitations to performance. Specifically, thermoelectric materials,
which are heavily doped semiconductors, require both high thermopower
and high conductivity. However, for a given carrier concentration
and temperature the thermopower increases with effective mass, while
the mobility (and conductivity) decreases. 

Here we elucidate a different effect of non-parabolicity that
can overcome this difficulty.
Within a parabolic band picture, the isoenergy surfaces in the band
structure, which govern transport take the shape of ellipsoids. In large
Fermi surface metals and in cases where the bands are degenerate at
the band edge (e.g. the $t_{2g}$ derived conduction band minimum of
some transition metal oxides) more complex shapes are possible, and
this can be beneficial in overcoming the above conundrum.
\cite{usui,shirai2013mechanism}
Here we show that there can be large deviations from the ellipsoidal
shapes that are usually described by an effective mass tensor even
for relatively low carrier concentrations of relevance to
thermoelectrics. This is the case even without band degeneracy.
Furthermore, this electronic feature
may be beneficial for thermoelectric properties.
Specifically, we discuss two narrow gap semiconductors,
GeTe and SnTe, which have
conduction band structures that are
not only anisotropic (i.e. with different effective masses in different directions)
and non-parabolic,
but in fact also have a novel
corrugated structure in the doping range
near the band edges that appears to affect thermoelectric
performance. This cannot be captured within parabolic band models, even with an
anisotropic effective mass tensor.

We show by first principles calculations that the performance of these
$n$-type materials may potentially outstrip that of
the $n$-type lead chalcogenides PbTe and PbSe.
This is of considerable importance given that
the best $n$-type $ZT$ for these chalcogenides
has significantly lagged the best reported $p$-type values of $ZT$
(see for example Ref. \onlinecite{zhang}).
We contrast these two materials with PbTe and PbSe, which
do not show this corrugated structure.

\begin{figure}
\centering 
\includegraphics[width=1.05\columnwidth,angle=0]{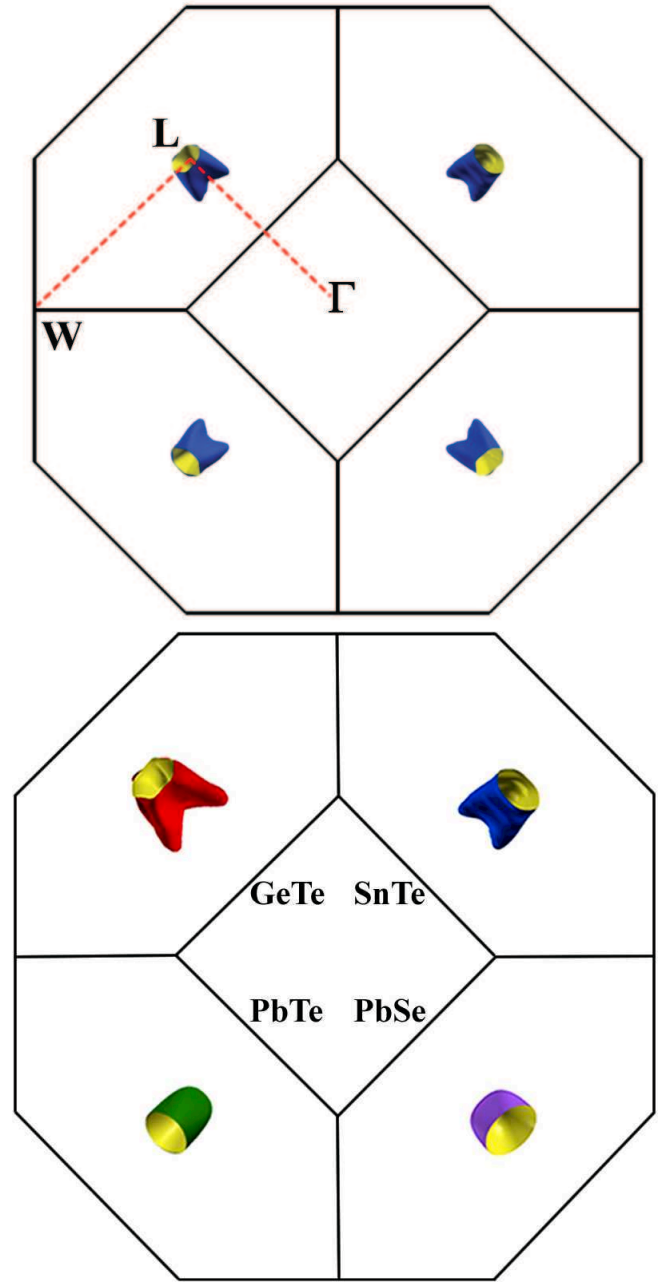}
\vspace{-3cm}
\caption{(color online) Calculated constant
energy surfaces near the conduction band minimum of SnTe (top panel).
Corresponding surfaces for GeTe, SnTe, PbTe, and PbSe
are shown in the four quadrants of the bottom panel.
The energy surfaces are plotted for a carrier concentration of
0.004 electrons per formula unit, which correspond to the electron concentration of 7.4 $\times10^{19}$, 6.3$\times 10^{19}$, 5.9$\times 10^{19}$, and 7.0 $\times10^{19}$ cm$^{-3}$ for GeTe, SnTe, PbTe, and PbSe, respectively. For the bottom panel the Fermi surfaces have been artificially enlarged, relative to the size of the
Brillouin zone, to better depict the corrugation.}
\label{isoenergy}
\end{figure}

We begin with the observation that many of the high-ZT materials,
such as PbTe, Bi$_2$Te$_3$ and PbSe
have highly anisotropic band structures.  
\cite{chung2000csbi4te6,mishra1997electronic,wang2007enhanced,takeuchi2004contribution,pei2011convergence} 
Typically, they have Fermi surfaces with rather different 
masses along different directions.
When the symmetry of the material is taken
into account, Fermi surface pocket degeneracy ensures that all directions
(in the case of the cubic rock-salt structure compounds
given above)
or all planar directions (for rhombohedral Bi$_2$Te$_3$) average
transport from both heavy and light bands. This leads to
both high thermopower and high electrical conductivity.
This thus represents one means of solving the conundrum of attaining
both these properties in the same bulk material.
The chalcogenides considered here have conduction band minima at the
$L$-point of the $fcc$ Brillouin zone.

The top panel of Fig. \ref{isoenergy} shows isoenergy surfaces for
SnTe near the conduction band minimum, while
the bottom panel of Fig. \ref{isoenergy}
shows a comparison of the four compounds, GeTe, SnTe, PbTe and PbSe.
These are based on 
calculations with augmented plane-wave plus local orbital (APW+lo)
method,
\cite{sjostedt2000alternative} 
as implemented in the WIEN2K code.\cite{wien2k}
We included spin-orbit coupling in our 
calculations and employed the modified Becke-Johnson potential of Tran and 
Blaha (TB-mBJ),
\cite{tran2009accurate}
which can generally yield accurate electronic band 
structures and gaps for simple semiconductors and insulators.
\cite{tran2009accurate,singh2010electronic,koller2011merits}
Within a parabolic band model, these constant energy surfaces
must take the shape of an ellipsoid of revolution about the
$\Gamma$-$L$ line, which the actual surfaces do for very low
carrier concentrations. 
It is clear from Fig. \ref{isoenergy} that the electron pockets
in SnTe have a corrugated shape, 
obviously different from that in PbTe and PbSe, where the pockets are relatively close 
to ellipsoids. This is even more noticeable in GeTe.  Note that there is still substantial Kane-band (i.e. quasilinear dispersion)
non-parabolicity in all four materials (as is evident in the band structure plots in Figure 2), but the corrugated character is absent in PbTe and PbSe.  

We note also from the Table that the difference in heavy and light masses ( a form of anisotropy) is greatest for SnTe and GeTe and significantly smaller for 
PbTe and PbSe.  This is also a contributor to improved transport in these materials, as the lighter band leads to good conductivity and the heavier
band to good thermopower, a scenario explored in Ref. \onlinecite{kuroki2007pudding}.

A simple argument (supported by calculations, below)
explains the benefits to thermoelectric transport of
the corrugated Fermi surfaces.
Briefly, the effect of the corrugation
is to increase the area of the Fermi surface
for a given carrier concentration
(the volume contained within the Fermi surface).
A larger area implies the presence of more electronic states,
or equivalently a larger electronic density of states,
which as is well known leads to larger thermopower for a given carrier
concentration.
Note, however, that this larger density-of-states is achieved
{\it without} a corresponding reduction in the electronic conductivity,
as occurs in the simple
case of an decrease in band dispersion, or equivalently an increase in
dispersive mass.
Instead the
{\it density of states} effective mass is increased by
the corrugated structure, but the {\it dispersive} mass
(i.e. $m^{* -1} = d^2 E_k/dk^2$) is not.
This is similar to, but distinct from, the ``heavy-band" effect in
the $p$-type lead chalcogenides,
which also increases the density-of-states without an increase in
dispersive mass.

We note that in a microscopic model, the scattering rate $\tau^{-1}(E)$
is sometimes taken as proportional to the electronic density-of-states (DOS),
so that in this scenario the increased DOS associated with the corrugation could
lead to increased scattering rates, hampering transport.  We have not attempted to quantify
or assess such an effect, but note that the aforementioned ``heavy-band" effect in the lead 
chalcogenides would in this scenario also lead to reduced transport, which does not occur - in fact,
the $p$-type lead chalcogenides show the highest $ZT$ values of {\it any} bulk material.  One additional
piece of information comes from computed values \cite{cohen} of the electron-phonon coupling constant $\lambda$
in $n$-type GeTe and SnTe, which are comparable and do not exceed 0.24 at any doping less than 2 $\times
10^{21}$cm$^{-3}$, a doping far heavier than optimal for these two materials (in fact, at optimal doping the
computed values fall between 0.1 and 0.18).  These relatively low values of the electron-phonon
coupling for GeTe and SnTe are strong evidence that good electronic conduction will in fact occur in 
these materials.

\vspace{0.5cm}
\hspace{8cm} {\bf Results}
\\
\\
For an isotropic parabolic band, the conductivity can be expressed in terms 
of the effective mass $m^{\ast }$ and the carrier concentration
$n$, $\sigma = ne^2\tau / m^\ast $.
Since the electrical conductivity intrinsically 
involves a scattering time $\tau(E)\sim E^{r}$, where $r$ is the scattering 
parameter, here we adopt two different approximations on $\tau $. 
One is CSTA ($r $= 0),
as widely used in \textit{ab-initio} calculations, including those in this paper.
In the other approximation, $\tau (E)$ is 
considered to be affected by the acoustic-mode lattice scattering ($r$ = -1/2), 
as considered in
Refs. \onlinecite{kim2009influence,cornett2012effect,neophytou2011effects,ParkerDavid2013-prl},
which we here note as ASTA. Then, considering 
the band degeneracy of 8 for our cases here (2 for spin and 4 coming from 
the number of electron pockets), we can derive the following expressions for 
thermopower within CSTA and ASTA according to the Mott formula \cite{snyder2008},
$S = {\frac{\pi ^2k_B^2 T}{3e}}{\frac{d\log [\sigma (E)]}{dE}}|_{E=E_F } $,
where $E_F$ is the Fermi energy,

\begin{equation}
\label{eq3}
S_{CSTA} = \frac{4\pi ^2k_B^2 }{eh^2}m^\ast T\left( {\frac{4\pi }{3n}} 
\right)^{2 / 3}
\end{equation}

\begin{equation}
\label{eq4}
S_{ASTA} = \frac{8\pi ^2k_B^2 }{3eh^2}m^\ast T\left( {\frac{4\pi }{3n}} 
\right)^{2 / 3}
\end{equation}

\begin{table}
\caption{\label{tab:table1}
Effective mass for lowest conduction band of GeTe, SnTe, PbTe, and 
PbSe (in m$_{0}$ units) along transverse ($W$-$L$) and longitudinal
($\Gamma$-$L$) directions.}
\begin{ruledtabular}
\begin{tabular}{ccccc}
{}& GeTe & SnTe&PbTe& PbSe\\ 
\hline
$m_T$ &0.022 & 0.021 & 0.040 &0.081\\
$m_L$ &0.49 & 0.18 &0.32 &0.18 \\
\end{tabular}
\end{ruledtabular}
\end{table} 

It is clear that with parabolic bands the thermopower is proportional
to $m^\ast$ for both approximations.  Note that the main effect of the ASTA is a reduction in thermopower
relative to the CSTA.  For the comparison of SnTe and GeTe with PbTe and PbSe we adopt the CSTA for all materials for consistency and
note that the likely effect of using ASTA would simply be a reduction in calculated thermopower for all four materials.  As will be seen, the likely advantages
of SnTe and GeTe may well dwarf any potential effects of neglecting the relaxation time energy dependence.

Note also that the Mott relation is inherently
a single band formula and neglects the effects of bipolar conduction.  However, as is the case here,
the region of likely optimal doping (i.e., likely largest $ZT$) is generally at concentrations
heavier than that where bipolar conduction becomes significant (i.e. to the right
of the thermopower maxima in Figure 4), where the single band Mott formula generally applies.  
To ensure a fair comparison with the specific 
compounds examined here, we also calculate the effective masses for the 
lowest conduction bands of GeTe, SnTe, PbTe and PbSe,
as shown in Table \ref{tab:table1}.   These masses were calculated
from a parabolic fitting of the dispersion very near the conduction band
minima, with axes chosen as the principal axes of the ellipsoids
(which run perpendicular to the Brillouin zone face around the L point).

We note that the effective mass of GeTe is the most different along 
transverse ($W$--$L$) and longitudinal directions ($\Gamma$--$L$),
as reflected in the
large anisotropy seen in its electron pockets
(see Fig. \ref{isoenergy}).
SnTe also shows very anisotropic, non-ellipsoidal pockets
although the difference is not as large as in GeTe.
Therefore, we chose the heavy effective mass of SnTe (0.18 $m_0$, 
with $m_0$ the free electron mass) in the parabolic model calculations. 

These differences can be seen more directly in Figure 2, which plots the
band structures and densities-of-states of the four compounds.  The conduction band DOS
rises from the band edge much more rapidly for SnTe and GeTe than the other two compounds, indicative
of the larger effective DOS mass of these two compounds.   This is responsible for the larger and comparable thermopower in these materials, irrespective
of the large mass of GeTe relative to SnTe (at 700 K the thermopower samples electronic structures larger than the energy window within which effective masses were determined).
Correspondingly, the conduction band for these two compounds shows
only a moderate dispersion of 0.9 eV from the L to W points, while the value is roughly double this for the other two compounds.  In addition, SnTe and GeTe
show near-degeneracies at the band edge, with at least one additional band minima only 0.15 eV above the CBM in GeTe and 0.4 eV above in SnTe.  These 
degeneracy differences, combined with the fact that the light-mass (as shown in the Table) for these two compounds is lower than for PbSe and PbTe, suggest that the conductivity
of SnTe and GeTe is likely to be substantial.  Indeed this is the case, as Figure 3 depicts the calculated values of $\sigma/\tau$ for all compounds, showing in
fact a slight advantage for SnTe and GeTe.

\begin{figure}
\centering 
\includegraphics[width=4.5in]{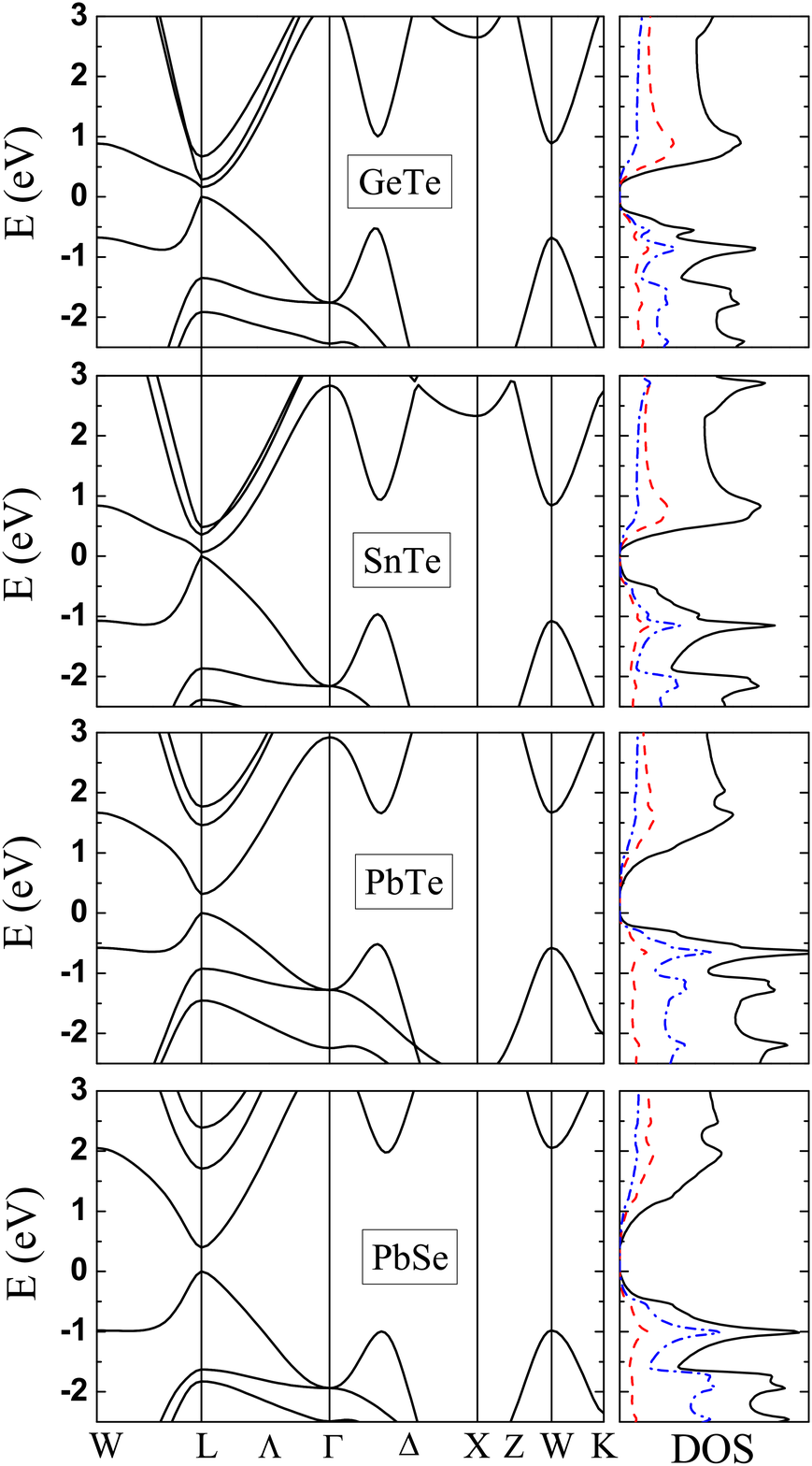}
\vspace{-3cm}
\caption{(color online) Calculated band structures and densities-of-states for
GeTe, SnTe, PbTe, and PbSe. }
\end{figure}

\begin{figure}
\centering 
\includegraphics[width=4.5in]{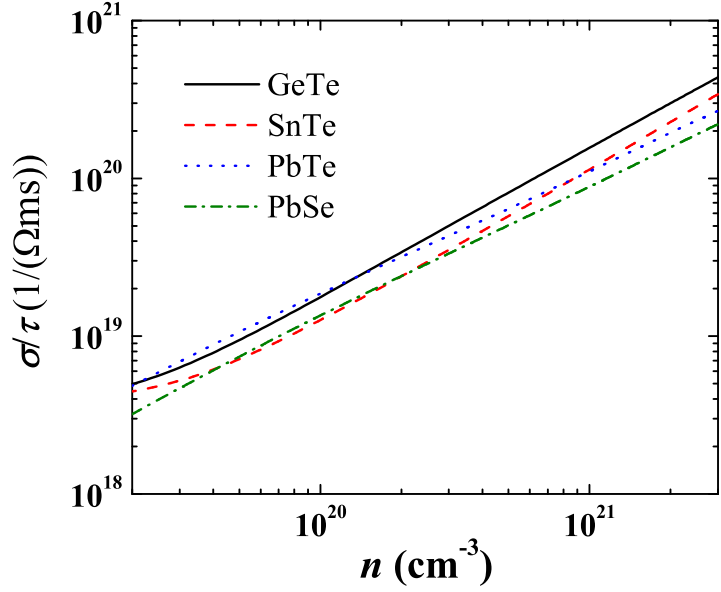}
\caption{(color online) Calculated $\sigma/\tau$ at 700 K for 
GeTe, SnTe, PbTe, and PbSe. }
\end{figure}

\begin{figure}
\centering 
\includegraphics[width=4.5in]{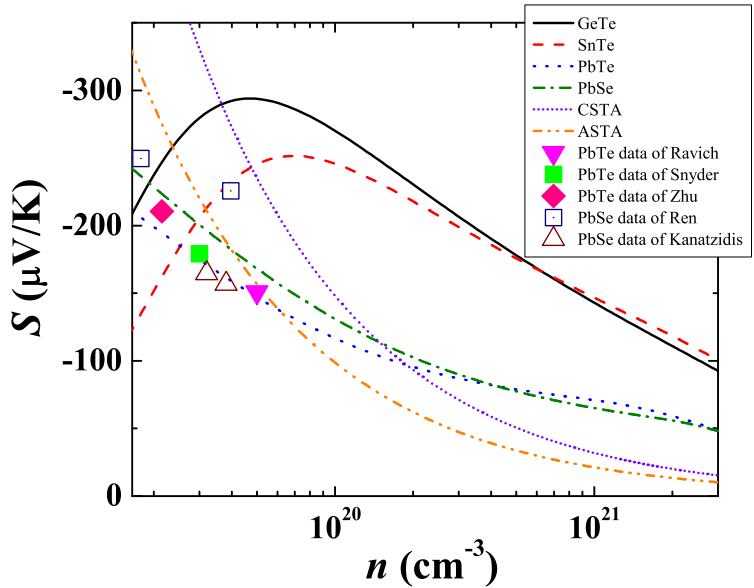}
\caption{\label{fig2}(color online) Calculated 700 K thermopower for
$n$-type GeTe, SnTe, PbTe, PbSe, and
general isotropic parabolic system within CSTA and ASTA, compared with the experimental data from Refs. \onlinecite{androulakis1,ravich,zhu1,pei1,zhang1}.  Note the excellent agreement
between theory and experiment for PbTe and the good agreement ($\pm 15$ percent) for PbSe.}
\end{figure}

Figure \ref{fig2} shows the thermopower, calculated within semiclassical Boltzmann
transport theory and the CSTA (see the Methods section)  at $T$ = 700 K.
We have chosen this temperature as it is above the rhombohedral
to cubic phase transition temperature of 670 K for GeTe, and is also a typical 
temperature for high temperature thermoelectric work.  While for applications such 
phase transitions can be a negative, in this work we are studying thermoelectric
performance irrespective of any application.  Note that our first principles calculations
follow fairly closely the existing experimental data on PbTe and PbSe, indicating the validity of our approach.

It is clear
that the behavior of $S$ for $n$-type GeTe and SnTe is significantly 
superior to that of PbTe and PbSe consistent with
previous numerical results.
\cite{xu2011thermoelectric,singh2010thermopower}
The enhanced thermopower in $n$-type GeTe and SnTe
comes from the corrugated Fermi surface character that increases
the Fermi surface area so that
more electronic states contribute to transport.
Furthermore, the thermopower of GeTe and SnTe obviously 
exceeds the parabolic band model
values obtained within CSTA at doping levels 
larger than 3.6$\times$10$^{19} $ and 4.8$\times$10$^{19}$ cm$^{-3}$,
for the two compounds,
respectively. This is the range where the properties are consistent
with good thermoelectric performance.
The beneficial effect of the non-parabolic
band structure is increased in the ASTA.
($S_{CSTA}$ is one and a half times larger than 
$S_{ASTA}$ according to the Eqs. (\ref{eq3}) and (\ref{eq4})).
The lead salts PbTe and PbSe
also have a larger thermopower than the parabolic case within ASTA at 
doping levels of $>$ 5.9$\times$10$^{19}$ and 3.9$\times$10$^{19}$ cm$^{-3}$,
respectively.

\begin{figure}
\includegraphics[width=4.5in]{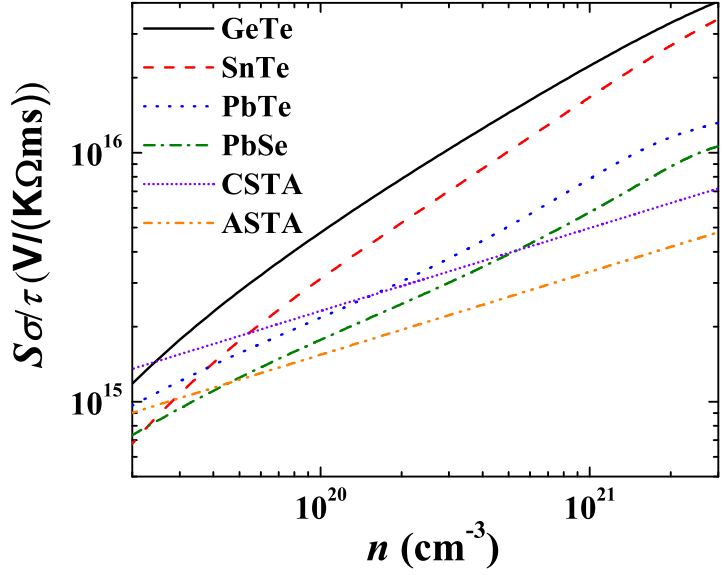}
\centering 
\caption{\label{fig3}(color online)
Calculated product of thermopower and conductivity
($S\sigma /\tau $) for n-type GeTe, SnTe, PbTe, PbSe,
and model isotropic parabolic band system within CSTA and ASTA at 700 K.
S is taken as the absolute value.}
\end{figure}

It is instructive, in order to isolate the effect of the corrugation, to consider the transport quantity
$S\sigma/\tau$ (Fig. \ref{fig3}),
as the effective mass in the
expressions of $\sigma $ and $S$ cancel in this product.
It is clearly seen from Fig. \ref{fig3} GeTe and SnTe have values
that substantially exceed those of the general parabolic case 
over a wide range of carrier concentration.  This appears to be due to the corrugation.  
PbTe and PbSe show higher
$S\sigma /\tau $ than the parabolic case
at a doping level of the order 10$^{20}$ cm$^{-3}$.
Remembering the larger $S$ in the specific anisotropic 
non-parabolic compounds, they have a much enhanced thermoelectric 
performance (proportional to power factor $S^{2}\sigma$)
than those of the parabolic 
case although actual values clearly depend on the details of the scattering.

Since for the actual quantiitative performance what is most critical is the power
factor $S^{2}\sigma$, rather than $S\sigma$, in Figure 6 we present a plot of the power factor
(with respect to scattering time $\tau$) S$^{2}\sigma/\tau$ for each of the four materials
for a range of carrier concentration.  While mean scattering times can and do vary from material to material,
such a difference is likely overshadowed by the band structure effects depicted in Figure 4, which
shows the substantial advantage of SnTe and GeTe relative to PbTe and PbSe. At the heavy dopings 
around $n=2-3 \times 10^{20}$ cm$^{-3}$, where thermoelectric performance is likely optimal for SnTe and GeTe, the power
factor is approximately {\it seven} times as large as for the other two materials.  While it is true that the relaxation times may differ, we take this plot
as strong evidence for good thermoelectric performance in SnTe and GeTe.

\begin{figure}
\includegraphics[width=4.5in]{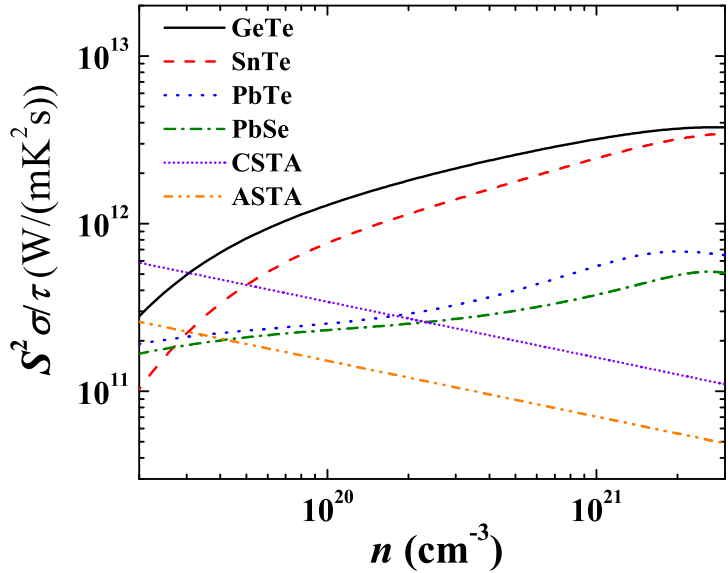}
\centering 
\caption{(color online)
Calculated product S$^2\sigma/\tau$  for n-type GeTe, SnTe, PbTe, PbSe,
and model isotropic parabolic band system within CSTA and ASTA at 700 K.}
\end{figure}
\vspace{0.5cm}
\hspace{8cm} {\bf Discussion}
\\
\\
In evaluating the transport coefficients for an isotropic parabolic band as 
above, the effective mass has a direct relationship with the carrier 
concentration $n$, which is dependent on the volume
of the Fermi surface
$4 \pi k_F^{3}/3$, $k_F=\frac{1}{\hbar}(2mE_F)^{1/2}$.
For the anisotropic case,
the effective mass is a second-rank tensor. $\sigma$ in the 
light mass direction at a given $E_{F}$ increases relative
to the isotropic case because the carrier 
concentration $n$ will be larger than that of the isotropic system with light 
mass in all directions. Accordingly, the conductivity of the anisotropic 
case will be higher if the scattering rate is not proportionately enhanced. 
On the other hand, $S(T)$, which is proportional to
$T/E_{F}$ at low temperature, 
the anisotropic system will have the same value as the isotropic light band 
case with the same Fermi energy. Conversely, for the given carrier 
concentration $n$, the anisotropic system will have a lower $E_{F}$ than the 
isotropic light band case, and therefore an increased $S(T)$.
Based on the above 
discussion, the anisotropic case will have a better 
thermoelectric performance than the isotropic light band system, supposing 
the scattering is not changed a lot. 

Now we turn to the isotropic system with a heavy band, just as the case we 
discussed in this work. The isotropic heavy band system could have an 
enhanced thermopower compared to
the corrugated anisotropic case because of a lower $E_{F}$ for 
the same carrier concentration. But this is only true at the lower carrier 
densities as shown in Fig. \ref{fig2}.
The dependence of $S$ on the carrier concentration
is much weaker than the -(2/3) power law for the
parabolic case. This means that with increasing carrier 
concentration, the thermopower in the corrugated case decreases much
slower 
than for the isotropic case. Therefore, anisotropic corrugated case will have an 
enhanced $S$ at high doping levels.
Taking GeTe as an example, at
carrier concentrations larger than 2.6$\times$10$^{20} $cm$^{-3}$,
the thermopower of GeTe exceeds that of the general isotropic case with the
heavy band mass of 0.49 $m_0$ using CSTA (not shown).
On the other hand,
it is obvious that anisotropic case will have a larger electronic 
velocity, and thus a higher conductivity than the isotropic system with 
heavy mass in all directions.
As illustrated in Fig. \ref{fig3}, GeTe has a superior 
$S\sigma /\tau $ over the isotropic system at
almost the entire carrier density ranges, which 
implies the higher conductivity in GeTe.
Specifically, the logarithmic increase rate in GeTe over
the range shown is $\sim$0.7, i.e. roughly
double the parabolic value of 1/3.
Thus, in anisotropic systems of this type the linkage
of thermopower and electrical 
conductivity is broken so that one may have effectively
heavy carriers that contribute 
to large $S$, and at the same time light carriers that provide high $\sigma$
in a single band system.
Related arguments have been made in the context
of oxides Na$_{x}$CoO$_{2}$, SrTiO$_{3}$,
and KTaO$_{3}$,\cite{kuroki2007pudding,usui,shirai2013mechanism,fan}
with complex multiband conduction band minima from degenerate $t_{2g}$
bands or a highly two dimensional metallic band structure as in
Na$_x$CoO$_2$. \cite{xiang-nac}
Here we find that such effects are important even
with a singly degenerate band
near the conduction band minimum of SnTe and GeTe.

In conclusion, we find that anisotropy
beyond the level possible in parabolic, ellipsoidal band models may
strongly influence transport properties even for moderately
doped semiconductors with finite gaps,
and that this
appears to be favorable for thermoelectric performance.
This is illustrated by detailed calculations for
GeTe and SnTe.  Based on our Boltzmann transport calculations, the $n$-type performance of these materials
may exceed that of the lead chalcogenides PbTe and PbSe.  We hope that our current findings 
will stimulate future experimental exploration
of the thermoelectric properties of these materials, particularly
$n$-type SnTe and alloys.

\vspace{0.8cm}
\hspace{8cm} {\bf Methods}
\\
\\
Transport properties of a crystalline solid, including the thermopower and 
conductivity $\sigma/\tau$ (here $\tau$ is an average scattering time) are calculated within 
Boltzmann transport theory,
based on the electronic structure.\cite{madsen2006boltztrap}
We performed electronic structure calculations using 
full-potential APW+lo method, using WIEN2K \cite{wien2k} including spin-orbit coupling.
To obtain reliable transport properties, a dense
48$\times$48$\times$48 $k$-point mesh
in the Brillouin zone was used.
The expressions for the electrical conductivity and the 
thermopower are:

\begin{equation}
\label{eq1}
S(T) = \frac{1}{\sigma (T)}\int {\sigma (E)(E - \mu ) \left( { - 
\frac{\partial f}{\partial E}} \right)dE}
\end{equation}

\begin{equation}
\label{eq2}
\sigma (T) =
\int {\sigma (E) \left( { - \frac{\partial f}{\partial E}} \right)dE}
\end{equation}

\noindent
where $f$ is the Fermi function,
\textit{$\mu $} is the chemical potential, and $\sigma $(E) 
is the transport function given by $N(E)\upsilon ^2(E)\tau (E)$.
The transport coefficients are tensors, but we omit the 
subscripts in the present work because of the cubic symmetry.
Calculations based on these formulas with actual band structures
have proven useful in rationalizing and 
predicting transport properties of known compounds.
\cite{scheidemantel,bertini,lykke,hamada,parker,wang2007enhanced,chen-cu2o,parker2010high,madsen2003electronic}
Here we used the BoltzTraP code.
\cite{madsen2006boltztrap}
The conductivity can 
only be calculated with respect to the relaxation time using the 
preceding expressions.
However, within the 
constant scattering time approximation (CSTA)
the thermopower can be obtained directly 
from the band structure information without adjustable parameters.
This approximation consists of assuming that the energy dependence of the
relaxation time at a given doping level
and temperature is negligible on the scale of
$k_BT$, and has shown considerable success in describing the thermopower
of a large number of materials.

\vspace{1.4cm}
\hspace{6cm} {\bf Author Contribution Statement}
\\
\\
X.C., D.P. and D.J.S. wrote the manuscript.  X.C. prepared figures 1-6.  All authors reviewed the manuscript.

\vspace{1cm}
\hspace{6.5cm} {\bf Additional Information}
\\
\\
The authors declare no competing financial interests.

\vspace{1cm}
\hspace{7cm} {\bf Acknowledgments}
\\
\\
This work was supported by the Department of Energy,
Basic Energy Sciences,
through the S3TEC Energy Frontier Research Center.

\end{document}